\documentstyle[12pt]{article}

\def\gr{\raise.3ex\hbox{$>$\kern-.75em\lower1ex\hbox{$\sim$}}}
\def\le{\raise.3ex\hbox{$<$\kern-.75em\lower1ex\hbox{$\sim$}}}
\mathchardef\Lag="724C
\begin{document}

\def\thefootnote{\fnsymbol{footnote}}

{\it University of Shizuoka}

\hspace*{10cm} {\bf US-94-05}\\[-.3in]

\hspace*{10cm} {\bf October 1994}\\[.4in]

\begin{center}

{\large\bf Table of Running Quark Mass Values : 1994}\\[.3in]

{\bf Yoshio Koide\footnote{
E-mail address: koide@u-shizuoka-ken.ac.jp}} \\[.1in]

Department of Physics, University of Shizuoka \\[.1in]
52-1 Yada, Shizuoka 422, Japan \\[.5in]

{\large\bf Abstract}\\[.1in]
\end{center}

\begin{quotation}
Running quark mass values $m_q(\mu)$ at some typical energy scales
($\mu=1$ GeV, $\mu=m_W$, and so on) are reviewed. The values depend
considerably on the value of $\Lambda_{\overline{MS}}$, especially,
the value of top quark mass at $\mu=1$ GeV does so. The relative ratios
of light quark masses ($m_u$, $m_d$ and $m_s$) to heavy quark masses
($m_c$, $m_b$ and $m_t$) are still controversial.
\end{quotation}

\newpage

\noindent {\bf \S 1. Introduction}

\vglue.1in

Recently, there has been considerable interest in phenomenological
studies of quark and lepton mass matrices in order to obtain a clue
to unified understanding of quarks and leptons.
However, for this purpose, we must have the reliable knowledge of
running quark mass values $m_q(\mu)$ which are evolved to an identical
energy scale $\mu$ (e.g. $\mu=1$ GeV). Since the earlier work by Gasser and
Leutwyler [1], many works [2-5] on estimates of running
quark masses have been
reported. However, the values of $\Lambda_{\overline{MS}}$
which were adopted in these references [2-5]
are not identical. Some of the input data have become
older. On the other hand, this year (1994), the first observation [6] of top
quark mass value has been reported, and the 1994 version of ^^ ^^ Review
of Particle Properties" (RPP94) [7] has been published.
Therefore, this year is just timely for summarizing these works at present
stage, and the review will be useful for physicists who intend to make
a model-building of quarks and leptons.

In this review, we will give a summary table of running quark masses
$m_q(\mu)$ at $\mu=1$ GeV, $\mu=m_q$, $\mu=m_W$ and $\mu=\Lambda_W$,
where $\mu=\Lambda_W$ is a symmetry breaking energy scale of
the electroweak gauge symmetry SU(2)$_L \times$U(1)$_Y$.
$$\Lambda_W\equiv \langle\phi^0\rangle=(\sqrt{2}G_F)^{-\frac{1}{2}}/\sqrt{2}=
174\ {\rm GeV} \ . \eqno(1.1)$$

In this paper, we use the mass renormalization equation
$$\mu\frac{d}{d{\mu}}m_q(\mu)=-\gamma(\alpha_s)m_q(\mu) \ , \eqno(1.2)$$
and do not use the renormalization equations for Yukawa couplings.
This prescription is applicable only to the  energy scale which is
below the symmetry breaking energy scale $\Lambda_W$ of the electroweak gauge
symmetry SU(2)$_L \times$U(1)$_Y$.
If we want to evolve our results $m_q(\mu)$ to an extremely high energy
scale far from $\mu=\Lambda_W$ (e.g. $\mu=\Lambda_{GUT}$), we must use the
renormalization equations for Yukawa couplings.

In the next section, we review  values of light quark masses
$m_u(\mu)$, $m_d(\mu)$ and $m_s(\mu)$ at $\mu=1$ GeV.
In \S 3, we review values of heavy quark masses $m_c(\mu)$,
$m_b(\mu)$ and $m_t(\mu)$ at $\mu=m_q$. In order to estimate
$m_q(\mu)$ at any $\mu$, we must know the values of the QCD
parameters $\Lambda^{(n)}_{\overline{MS}}$ (n=3,4,5,6).
In \S 4, the values of $\Lambda^{(n)}_{\overline{MS}}$ are evaluated.
In \S 5, the values of $m_q(\mu)$ at $\mu=1$ GeV, $\mu=m_q$, $\mu=m_w$
and $\mu=\Lambda_W$ are estimated.
Finally, \S 6 is devoted to summary and discussion.

\vspace*{.2in}

\noindent {\bf \S 2. Light quark masses}

\vglue.1in

Grasser and Leutwyler [1] have concluded in their review article of
1982 that the light quark masses $m_u(\mu)$, $m_d(\mu)$ and $m_s(\mu)$
at $\mu=1$ GeV
are
$$
\begin{array}{lrrl}
m_u(1{\rm GeV})= & 5.1 & \pm 1.5 & {\rm MeV} \ , \\
m_d(1{\rm GeV})= & 8.9 & \pm 2.6 & {\rm MeV} \ , \\
m_s(1{\rm GeV})= & 175 & \pm 55  & {\rm MeV} \ .
\end{array}\eqno(2.1)
$$

On 1987, Domingues and Rafael [2] have re-estimated those values.
They have obtained the same ratios of the light quark masses
with those estimated by Grasser and Leutwyler,
but they have used a new value of ($m_u+m_d$) at $\mu=1$ GeV
$$
\begin{array}{lrrl}
(m_u+m_d)_{\mu=1GeV}= & (15.5 & \pm 2.0) & {\rm MeV} \ ,
\end{array}\eqno(2.2)
$$
instead of Grasser--Leutwyler's value $(m_u+m_d)_{\mu=1GeV}=(14\pm4)$
MeV.
Therefore, Dominguez and Rafael have concluded as
$$
\begin{array}{lrrl}
m_u(1{\rm GeV})= & 5.6 & \pm 1.1 & {\rm MeV} \ , \\
m_d(1{\rm GeV})= & 9.9 & \pm 1.1 & {\rm MeV} \ , \\
m_s(1{\rm GeV})= & 199 & \pm 33  & {\rm MeV} \ .
\end{array}\eqno(2.3)
$$

Narison (1989) [3] has obtained
$$
\begin{array}{lrrl}
m_u(1{\rm GeV})= & 5.2 & \pm 0.5 & {\rm MeV} \ , \\
m_d(1{\rm GeV})= & 9.2 & \pm 0.5 & {\rm MeV} \ , \\
m_s(1{\rm GeV})= & 159.5 & \pm 8.8 & {\rm MeV} \ ,
\end{array}\eqno(2.4)
$$
by using $(m_u+m_d)_{\mu=1\ {\rm GeV}}=(14.4 \pm 1.0)$ MeV.

On the other hand, Donoghue and Holstein (1992) [4] have estimated somewhat
different quark mass rations
$$
\begin{array}{l}
r_1= (m_u+m_d)/ [m_s+(m_u+m_d)/2]= 0.061 \ , \\
r_2= (m_d-m_u)/ [m_s-(m_u+m_d)/2]= 0.036 \ .
\end{array}\eqno(2.5)
$$
which lead to
$$
\begin{array}{l}
m_d/m_u= 3.49 , \ \ \ m_s/m_d= 20.7 \ .
\end{array}\eqno(2.6)
$$
The value of $m_d/m_u$ is considerably different from the previous values,
e.g.,  Grasser--Leutwyler's  value $m_d/m_u=1.75$.
Donoghue and Holstein estimated the values (2.5)
from the following four different sources:
(1) $r_1r_2=2.11\times10^{-3}$ from meson masses $+(\Delta m_R^2)_{EM}$ ,
(2) $r_1r_2=2.35\times10^{-3}$ from $\eta\rightarrow 3\pi$ decay ,
(3) $r_1/r_2=0.67\pm0.16$ from $\psi^{'}\rightarrow J/\psi+\pi^0(\eta)$ , and
(4) $r_1=0.067\pm0.012$ from meson masses and $L_7$ .
The values of $r_1$ and $r_2$ from these sources are still controversial.

Donoghue and Holstain's value of $m_s/m_d$ is in good agreement with
that estimated by Dominquez and Rafael.
Hereafter, we will adopt Dominguez-Rafael's value (2.3)
as light quark mass values at $\mu=1$ GeV.

\vglue.2in


\noindent {\bf \S 3. Heavy quark masses at $\mu=m_q$}

\vglue.1in

\noindent \underline{Pole mass}

Sometimes, values of heavy quark masses $m_c$, $m_b$, and $m_t$ are
estimated in terms of the ^^ ^^ pole" masses $M_q^{\rm pole}$.
It is known that the pole mass, $M_q^{\rm pole}(p^2=m_q^2)$, is
a gauge-invariant, infrared-finite, renomalization-scheme-independent
quantity.

Generally, mass function $M(p^2)$, which is defined by [1]
$$S(p)=Z(p^2)/\left(M(p^2)-\not\!p \right) \ , \eqno(3.1)$$

$$Z(p^2)=1-\frac{\alpha_s}{3\pi}(a-3b+\frac{2}{3})\lambda+O(\alpha_s^2) \ ,
\eqno(3.2)$$
is related to
$$M(p^2)=m(\mu)\left[1+\frac{\alpha_s}{\pi}(a+\lambda b)+O(\alpha_s^2)
\right ] \ ,
\eqno(3.3)$$
$$
 a=\frac{4}{3}-\ln\frac{m^2}{\mu^2}+\frac{m^2-p^2}{p^2}
\ln\frac{m^2-p^2}{m^2} \ , \eqno(3.4)$$
$$
b=-\frac{m^2-p^2}{3p^2}\left(1+\frac{m^2}{p^2}
\ln\frac{m^2-p^2}{m^2}\right) \ , \eqno(3.5)$$
where $\lambda$ is given by $\lambda=0$ in the Landau gauge and $\lambda=1$
in the Feynman gauge.
For $p^2=m^2$, we obtain $a=4/3$ and $b=0$, so that we obtain
the relation
$$M_q^{\rm pole}(p^2=m_q^2)=m_q(m_q)\left(1+\frac{4}{3}\frac{\alpha_s}{\pi}+
O(\alpha_s^2)\right) \ . \eqno(3.6)$$

The estimate of the pole mass to two loops has been given by
Gray {\it et al} [8]:
$$
m_q(m_q)=M_q^{pole}(p^2=m_q^2)\left[1
-\frac{4}{3}\frac{\alpha_s(M_q^{pole})}{\pi}-
\left(K-\frac{16}{9}\right)\left(\frac{\alpha_s(M_q^{pole})}{\pi}\right)^2
+O(\alpha_s^3)\right] \ , \eqno(3.7)$$
$$
K = K_0 +\frac{4}{3}\sum_{i=1}^{n-1} \Delta(M_i^{pole}/M_n^{pole})
\simeq 17.15 -1.04 n + \frac{4}{3}\times 1.04\sum_{i=1}^{n-1}
\frac{M_{i}^{pole}}{M_q^{pole}}
\ . \eqno(3.8)$$
Here the sum in (3.8) is taken over $n-1$ light quarks with masses
$M_i^{pole}$ ($M_{i}^{pole}<M_{n}^{pole}\equiv M_q^{pole}$).
The exact expressions of $K_0$ and $\Delta(r)$ are given in Ref.~[8].
The numerical values of  $\Delta(M_i^{pole}/M_n^{pole})$ without
approximation are tabled in Surguladze's paper [9]

Similarly, for the spacelike value of $p^2$, $p^2=-m_q^2$,
we obtain $a=4/3-2\ln2$ and $b=(2/3)(1-\ln 2)$, so that we obtain
the gauge-dependent ``Euclidean" masses
$$M_q^{pole}(p^2=-m_q^2)=m_q(m_q)\left[1+\frac{\alpha_s}{\pi}(\frac{4}{3}
-2\ln2)+O(\alpha_s^2)\right] \ . \eqno(3.9)$$

\vglue.1in
\noindent \underline{Charm and bottom quark masses}

Gasser and Leutwyler (1982) [1] have estimated charm and bottom quark masses
$m_c$ and $m_b$ as
$$m_c(m_c)=1.27\pm0.05 \ {\rm GeV} \ , \eqno(3.10)$$
$$m_b(m_b)=4.25\pm0.10 \ {\rm GeV} \ . \eqno(3.11)$$

Narison (1989) [3] has, from $\psi$- and $\Upsilon$-sum rules,
estimated those as
$$M_c^{pole}(p^2=-m_c^2)=1.26 \pm 0.02 \ {\rm GeV} \ , \eqno(3.12)$$
$$M_b^{pole}(p^2=-m_b^2)=4.23 \pm 0.05 \ {\rm GeV} \ , \eqno(3.13)$$
which mean
$$M_c^{pole}(p^2=m_c^2)=1.45 \pm 0.05 \ {\rm GeV} \ , \eqno(3.14)$$
$$M_b^{pole}(p^2=m_b^2)=4.67 \pm 0.10 \ {\rm GeV} \ , \eqno(3.15)$$
with $\Lambda=0.15 \pm 0.05$ GeV.

Dominguez and Paver (1992) [5] have estimated the value of $m_b$ as
$$M_b^{pole}(p^2=m_b^2)=4.72 \pm 0.05 \ {\rm GeV} \ , \eqno(3.16)$$
from the ratio of Laplace transform QCD sum rules in the non-relativistic
limit which is not so dependent on the value of $\Lambda$.

Recently, Tirard and Yudur\'{a}in [10] have re-estimated charm and
bottom quark masses precisely and rigorously.
They have concluded that
$$
M_c^{pole}(p^2=m_c^2)=1.570\pm 0.019 \mp 0.007 \ {\rm GeV}\ ,
\eqno(3.17)
$$
$$
M_b^{pole}(p^2=m_b^2)=4.906_{-0.051}^{+0.069} \mp 0.004_{-0.040}^{+0.011}
 \ {\rm GeV}\ , \eqno(3.18)
$$
$$
m_c(m_c)=1.306_{-0.034}^{+0.021}\pm 0.006 \ {\rm GeV} \ , \eqno(3.19)
$$
$$
m_b(m_b)=4.397_{-0.002+0.004-0.032}^{+0.007-0.003+0.016} \ {\rm GeV}
\ , \eqno(3.20)
$$
where the first- and second-errors come from the use of the QCD parameter
$\Lambda_{\overline{MS}}^{(4)}=0.20_{-0.06}^{+0.08}$ GeV and
the gluon condensate value $\langle\alpha_s G^2\rangle=0.042\pm 0.020$
GeV$^4$, and the third error denotes a systematic error.
They have used $K_c\simeq 14.0$ and $K_b\simeq 13.4$
as the values of $K_c$ and $K_b$ given by (3.8).

Hereafter, we adopt Tirard and Yudur\'{a}in's values (3.19)
and (3.20) as $m_c(m_c)$ and $m_b(m_b)$,
although we do not adopt their value $\Lambda_{\overline{MS}}^{(4)}=0.20$ GeV
as $\Lambda_{\overline{MS}}^{(n)}$.
For simplicity, we refer the values (3.19) and (3.20) as
$$
m_c(m_c)=1.306_{-0.035}^{+0.022} \ {\rm GeV} \ , \eqno(3.21)
$$
$$
m_b(m_b)=4.397_{-0.033}^{+0.018} \ {\rm GeV} \ . \eqno(3.22)
$$

\vglue.1in
\noindent \underline{Top quark mass}

Recently, the CDF collaboration (1994) [6] has reported
the top quark mass value
$$m_t=174 \pm 10_{-12}^{+13} \ {\rm GeV} \ \eqno(3.23)$$
from the data of $p\overline{p}$ collisions at $\sqrt{s}=1.8$ TeV.
The value (3.23) is consistent with the recent
standard-model-fitting value [11]
$$m_t=161^{+15+16}_{-16-22}\ {\rm GeV} \ , \eqno(3.24)$$
from LEP and $p\overline{p}$ collider data.

We adopt the value (3.23) as the top quark mass value at
$\mu=m_t$.
Hereafter, we will simply refer the value (3.23) as
$$
     m_t(m_t)=174^{+22}_{-27} \ {\rm GeV} \ . \eqno(3.25)
$$

Note that usually the so-called standard-model-fitting value of $m_t$
does not correspond to $m_t(m_t)$ but to $M_t^{pole}(p^2=m_t^2)$.
The CDF value of $m_t(m_t)$, (3.25), together with the value of
$\Lambda_{\overline{MS}}^{(5)}=0.195$ GeV [7] (see the next section),
leads to
$$
M_t^{pole}(p^2=m_t^2)=182_{-28}^{+23} \ {\rm GeV}\ . \eqno(3.26)
$$

\vspace*{.3in}


\noindent {\bf \S 4. Estimates of the values of
$\Lambda_{\overline{MS}}^{(n)}$}

\vglue.1in

Prior to estimates of the running quark masses $m_q(\mu)$, we must estimate
the values of $\Lambda_{\overline{MS}}^{(n)}$.

The effective QCD coupling $\alpha_s=g_s^2/4\pi$ is controlled by the
$\beta$-function:
$$\mu\frac{\partial\alpha_s}{\partial\mu}=\beta(\alpha_s) \ , \eqno(4.1)$$
where
$$\beta(\alpha_s)=-\frac{\beta_0}{2\pi}\alpha_s^2-\frac{\beta_1}{4\pi^2}
\alpha_s^3+O(\alpha_s^4) \ , \eqno(4.3)$$

$$\beta_0=11-\frac{2}{3}n_q , \ \ \ \beta_1=51-\frac{19}{3}n_q
\ , \eqno(4.4)$$
and $n_q$ is the effective number of quark flavors, so the $\alpha_s(\mu)$
is given by\footnote{
In RPP94 [7] a three-loop expression of $\alpha_s(\mu)$ has been reviewed.
However, at the moment, the two-loop expression (4.5) is sufficient
for estimating running quark mass values to two-loops. }
$$\alpha_s (\mu)=\frac{4\pi}{\beta_0}\frac{1}{L}\left[1-
\frac{2\beta_1}{\beta_0^2}\frac{\ln L}{L}+
O(L^{-2}\ln^2 L)\right] \ . \eqno(4.5)$$
where
$$L=\ln(\mu^2/\Lambda^2) \ . \eqno(4.6)$$
At present, we can use only the expression of $\alpha_s(\mu)$
where the higher order term $O$ in (4.5) is dropped. Then, the value of
$\alpha_s(\mu)$ is not continuous at $n$th quark threshold $\mu_n$ (at
which the $n$th quark flavor channel is opened), because the coefficients
$\beta_0$ and $\beta_1$ in (4.2) depend on the effective quark flavor
number $n_q$.
Therefore, usually, we use the expression $\alpha_s^{(n)}(\mu)$ (4.5)
with a different $\Lambda_{\overline{MS}}^{(n)}$ for each energy scale range
$\mu_n\leq\mu<\mu_{n+1}$, where $\Lambda_{\overline{MS}}^{(n)}$ are
defined such as $\Lambda_{\overline{MS}}^{(n-1)}$ and
$\Lambda_{\overline{MS}}^{(n)}$
satisfy the relation
$$\alpha_s^{(n-1)}(\mu_n)=\alpha_s^{(n)}(\mu_n) \ . \eqno(4.7)$$
Therefore, we practically regard $n$th quark mass value $m_{qn}(\mu)$
at $\mu=m_{qn}$, $m_{qn}(m_{qn})$, as $\mu_n$.

Particle data group (PDG) [7] has concluded that the world average value
of $\Lambda_{\overline{MS}}^{(5)}$ is
$$\Lambda_{\overline{MS}}^{(5)}=195^{+65}_{-50}
{\rm MeV} \ . \eqno(4.8)
$$
On the other hand, in the conventional quark mass estimates since Gasser-
Leutwyler [1], the value $\Lambda_{\overline{MS}}^{(3)}=150$ MeV is
frequently used, although the value was used in the one-loop
expression of $\alpha_s(\mu)$.
For reference, we estimate $\Lambda_{\overline{MS}}^{(n)}$ and $m_q(\mu)$
for the case of $\Lambda_{\overline{MS}}^{(3)}=150$ MeV as well as
the case of  $\Lambda_{\overline{MS}}^{(5)}=195$ MeV.

Starting from $\Lambda_{\overline{MS}}^{(5)}\equiv 0.195$ GeV,
by using the continuity condition of $\alpha_s(\mu)$, (4.7),
at $\mu_5=m_b(m_b)=4.397$ GeV, $\mu_4=m_c(m_c)=1.306$ GeV,
and  $\mu_6=m_t(m_t)=174$ GeV,
we obtain $\Lambda_{\overline{MS}}^{(4)}=0.28475$ GeV.
$\Lambda_{\overline{MS}}^{(3)}=0.33156$ GeV and
$\Lambda_{\overline{MS}}^{(6)}=0.07760$ GeV.
These results are summarized in Table IV.

Similarly, the values of $\Lambda_{\overline{MS}}^{(n)}$ are
estimated for the case of $\Lambda_{\overline{MS}}^{(3)}\equiv 0.150$ GeV.
The results are listed in Table IV.

\vglue.2in
\begin{minipage}[tl]{6cm}
Table IV. The values of $\Lambda_{\overline{MS}}^{(n)}$
in unit of GeV and $\alpha_s(\mu_n)$.

\noindent
The underlined values are input values.
Here, $\mu_4=m_c(m_c)=1.306$ GeV, $\mu_5=m_b(m_b)=4.397$ GeV,
$\mu_6=m_t(m_t)=174$ GeV, and $m_Z=91.187$ GeV are used.
\end{minipage}
\begin{minipage}[tr]{7cm}
$$
\begin{array}{|l|l|l|}\hline
 & \ \ \ {\rm Case\ I}\ \ \ & \ \ \ {\rm Case\ II}\ \ \  \\[.1in] \hline
\Lambda_{\overline{MS}}^{(3)} & 0.33156 & \underline{0.15000} \\ [.1in]
\Lambda_{\overline{MS}}^{(4)} & 0.28475 & 0.11585 \\ [.1in]
\Lambda_{\overline{MS}}^{(5)} & \underline{0.19500} & 0.07164 \\ [.1in]
\Lambda_{\overline{MS}}^{(6)} & 0.07760 & 0.02562 \\[.1in] \hline
\alpha_s(\mu_4) & 0.36122 & 0.23632 \\ [.1in]
\alpha_s(\mu_5) & 0.22122 & 0.16554 \\ [.1in]
\alpha_s(\mu_6) & 0.10539 & 0.09295 \\ [.1in]
\alpha_s(m_Z) & 0.11541 & 0.10606 \\[.1in] \hline
\end{array}
$$
\end{minipage}

\newpage

\noindent {\bf \S 5. Estimates of running quark masses}

\vglue.1in

The scale dependence of a running quark mass $\mu_q(\mu)$ is
determined by the equation
$$\mu\frac{d}{d{\mu}}m_q(\mu)=-\gamma(\alpha_s)m_q(\mu) \ , \eqno(5.1)$$
where
$$\gamma(\alpha_s)=\alpha_s\gamma_0+\alpha_s^2\gamma_1+O(\alpha_s^3)
\ , \eqno(5.2)$$
$$\gamma_0=2 \ , \ \ \ \gamma_1=\frac{101}{12}-\frac{5}{18}n_q \ ,
\eqno(5.3) $$
so that $m_q(\mu)$ is given by
$$m_q=\widetilde{m}_q\left(\frac{1}{2}L\right)^{-2\gamma_0/\beta_0}
\left[1-\frac{2\beta_1\gamma_0}{\beta_0^3}
\frac{\ln L+1}{L}+\frac{8\gamma_1}{\beta_0^{2}L}+O(L^{-2}\ln^2 L)\right] \ ,
\eqno(5.4)$$
where $\beta_0$ and $\beta_1$ are given in (4.3) and $L=\ln(\mu^2/\Lambda^2)$.
Here, $\widetilde{m}_q$ is the renormalization group invariant mass, which is
independent of $\ln(\mu^2/\Lambda^2)$.

Since we interest only in the rations $m_q(\mu)/\widetilde{m}_q$,
we define the following quantity
$$
R^{(n)}=\left(\frac{1}{2}L\right)^{-2\gamma_0/\beta_0}
\left(1-\frac{2\beta_1\gamma_0}{\beta_0^3}
\frac{\ln L+1}{L}+\frac{8\gamma_1}{\beta_0^{2}L}\right) \ . \eqno(5.5)$$
The value of $R^{(n)}$ is not continuous at
$\mu=\mu_n$ ($\mu_n$ is the $n$th quark flavor threshold). Therefore, we
calculate the evolution of the quark masses $m_q(\mu)$ from $\mu=\mu_A$ (
$\mu_m\leq\mu_A<\mu_{m+1}$) to $\mu=\mu_B$ ($\mu_n\leq\mu_B<\mu_{n+1}$)
as follows:
$$
\frac{m_q(\mu_B)}{m_q(\mu_A)}=\left(\frac{R^{(m)}(\mu_{m+1})}{R^{(m)}{(\mu_A)}}
\right) \left(\frac{R^{(m+1)}(\mu_{m+2})}{R^{(m+1)}(\mu_{m+1})}\right)
\cdots\left(\frac{R^{(n-1)}(\mu_n)}{R^{(n-1)}(\mu_{n-1})}\right)
\left(\frac{R^{(n)}(\mu_B)}{R^{(n)}(\mu_n)}\right) \ . \eqno(5.6)$$
For example, the ratio $m_t(m_W)/m_t$(1 GeV) is given by
$$
\frac{m_t(m_W)}{m_t(1 {\rm GeV})}=
\left(\frac{R^{(3)}(m_c)}{R^{(3)}(1 {\rm GeV})}\right)
\left(\frac{R^{(4)}(m_b)}{R^{(4)}(m_c)}\right)
\left(\frac{R^{(5)}(m_W)}{R^{(5)}(m_b)}\right)
\ . \eqno(5.7)
$$
The values of $R^{(4)}(m_b)/R^{(4)}(m_c)$,
$R^{(5)}(m_t)/R^{(5)}(m_b)$,
and so on are summarized in Table V.

\vglue.2in

\begin{quotation}
Table V. Values of $R^{(\mu)}(\mu)$ for the case I
($\Lambda^{(5)}_{\overline{MS}}$ = 0.195 GeV)
and  the case II
($\Lambda^{(3)}_{\overline{MS}}$ = 0.150 GeV).
\end{quotation}
$$
\begin{array}{|l|l|l|}\hline
 & \Lambda^{(5)}=0.195\ {\rm GeV} & \Lambda^{(3)}=0.150\ {\rm GeV} \\[.1in]
\hline
R^{(3)}(1 {\rm GeV}) & 1.00886 \ \ \ \equiv 1 & 0.738347 \ \ \ \equiv 1
\\[.1in]
R^{(3)}(m_c) & 0.882993 \ \ \ 0.87524 & 0.69043 \ \ \ 0.93510 \\[.1in] \hline
R^{(4)}(m_c) & 0.84169 \ \ \ \equiv 1 & 0.64410 \ \ \ \equiv 1 \\[.1in]
R^{(4)}(m_b) & 0.60363 \ \ \ 0.71716 & 0.52206 \ \ \ 0.81052 \\[.1in] \hline
R^{(5)}(m_b) & 0.55141 \ \ \ \equiv 1 & 0.47152 \ \ \ \equiv 1 \\[.1in]
R^{(5)}(m_W) & 0.38415 \ \ \ 0.69667 & 0.35418 \ \ \ 0.75115 \\[.1in]
R^{(5)}(m_t) & 0.36036 \ \ \ 0.65353 & 0.33529 \ \ \ 0.71111 \\[.1in] \hline
R^{(6)}(m_t) & 0.31051 \ \ \ \equiv 1  & 0.28700 \ \ \ \equiv 1 \\[.1in]
R^{(6)}(\Lambda_W) & 0.31051 \ \ \ 1.00000  & 0.28700 \ \ \ 1.00000 \\[.1in]
\hline
\end{array}
$$

In Table VI, we summarize the running quark mass values at $\mu=m_q$,
$\mu=1$ GeV, $\mu=m_W(=80.22$ GeV) and $\mu=\Lambda_W(=174$ GeV),
where $\Lambda_W$ is defined by
$$\Lambda_W\equiv \langle\phi^0\rangle=(\sqrt{2}G_F)^{-\frac{1}{2}}/\sqrt{2}=
174\ {\rm GeV} \ . \eqno(5.8)$$

\newpage

\begin{quotation}
Table VI. Running quark mass values $m_q(\mu)$ (in unit of GeV)
at $\mu=m_q$, $\mu=1$ GeV, $\mu=m_W=80.22$ GeV and $\mu=\Lambda_W=174$ GeV.
The upper values (lower values) are the running quark mass values in the case
of $\Lambda^{(5)} \equiv 0.195$ GeV (the case of $\Lambda^{(3)}
\equiv 0.150$ GeV).
\end{quotation}
$$
\begin{array}{|c|c|c|c|c|}\hline
 & m_q(m_q) & m_q(1 {\rm GeV}) & m_q(m_W) & m_q(\Lambda_W) \\[.1in] \hline
m_u & 0.3463^{+0.0017}_{-0.0018} & 0.0056 \pm 0.0011
& 0.00245\pm 0.00048 & 0.00230\pm 0.00045 \\[.1in]
 & (0.1631^{+0.0015}_{-0.0017}) & (0.0056 \pm 0.0011)
& (0.00319 \pm 0.00063) & (0.00302 \pm 0.00059) \\[.1in] \hline
m_d & 0.3524^{+0.00013}_{-0.0015} & 0.0099 \pm 0.0011
& 0.00433 \pm 0.00048 & 0.00406\pm 0.00045 \\[.1in]
 & (0.169 \pm 0.019) & (0.0099 \pm 0.0011)
& (0.00564 \pm 0.00063) & (0.00534 \pm 0.00059) \\[.1in] \hline
m_s & 0.489 \pm 0.021 & 0.199 \pm 0.033
& 0.087 \pm 0.014 & 0.082 \pm 0.014 \\[.1in]
 & (0.338 \pm 0.029) & (0.199 \pm 0.033)
& (0.113 \pm 0.019) & (0.107 \pm 0.018) \\[.1in] \hline
m_c & 1.306^{+0.022}_{-0.035} & 1.492^{+0.023}_{-0.040}
& 0.653^{+0.009}_{-0.017}  & 0.612^{+0.010}_{-0.016}  \\[.1in]
 & (1.306^{+0.022}_{-0.035}) & (1.397^{+0.024}_{-0.037})
& (0.795^{+0.013}_{-0.021}) & (0.753^{+0.013}_{-0.023}) \\[.1in] \hline
m_b & 4.397^{+0.018}_{-0.033} & 7.005^{+0.029}_{-0.053}
& 3.063^{+0.013}_{-0.023}  & 2.874^{+0.012}_{-0.022}  \\[.1in]
 & (4.397^{+0.018}_{-0.033}) & (5.801^{+0.024}_{-0.044})
& (3.303^{+0.014}_{0.025}) & (3.127^{+0.013}_{-0.023}) \\[.1in] \hline
m_t & 174^{+22}_{-27} & 424^{+54}_{-66} & 185^{+23}_{-29} & 174^{+22}_{-27}
\\[.1in]
 & (174^{+22}_{-27}) & (323 ^{+41}_{-50}) & (184^{+23}_{-29})
& (174^{+22}_{-27}) \\[.1in] \hline
\end{array}
$$

\vglue.3in


\noindent {\bf \S 6. Summary}

\vglue.1in

We have estimated running quark mass values $m_q(\mu)$ at $\mu=m_q$,
$\mu=1$ GeV, $\mu=m_W=80.22$ GeV and $\mu=\Lambda_W=174$ GeV
for the two cases,
$\Lambda^{(5)}=0.195$ GeV ($\Lambda^{(3)}=0.332$ GeV, $\Lambda^{(4)}=0.285$
GeV, $\Lambda^{(6)}=0.0776$ GeV) and $\Lambda^{(3)}=0.150$ GeV
($\Lambda^{(4)}=0.116$ GeV, $\Lambda^{(5)}=0.0716$ GeV,
$\Lambda^{(6)}=0.0256$ GeV).
Of course, the case of $\Lambda^{(3)}=0.150$ GeV has been listed only for
reference, it is not our conclusion.

We have adopted the following quark mass values as the input values:

\noindent for light quark masses, Dominuez-Rafael's values:
$$
\begin{array}{lrrl}
m_u(1{\rm GeV})= & 5.6 & \pm 1.1 & {\rm MeV} \ , \\
m_d(1{\rm GeV})= & 9.9 & \pm 1.1 & {\rm MeV} \ , \\
m_s(1{\rm GeV})= & 199 & \pm 33  & {\rm MeV} \ ,
\end{array}\eqno(2.3)
$$
for charm and bottom quarks, Tirard and Yudur\'{a}in's values:
$$
m_c(m_c)=1.306_{-0.035}^{+0.022} \ {\rm GeV} \ , \eqno(3.21)
$$
$$
m_b(m_b)=4.397_{-0.033}^{+0.018} \ {\rm GeV} \ . \eqno(3.22)
$$
and, for top quark mass,  CDF value:
$$m_t=174^{+22}_{-27}\ {\rm GeV} \ . \eqno(3.25)$$

The results are summarized in Table VI.
As seen in Table VI, the running quark mass values (especially, those of
heavy quarks at $\mu=1$ GeV, and those of light quarks at $\mu=m_W$ and
$\mu=\Lambda_W$) are highly dependent on the value of
$\Lambda_{\overline{MS}}$.
The value of $\Lambda_{\overline{MS}}$ given in (4.8) includes large
error values, so that the absolute values of quark masses in Table VI are
not conclusive.

Although in Table VI, the values of $m_q(m_q)$ for light quarks are
listed, those values, especially those for $u$ and $d$,
should not be taken rigidly, because
$\alpha_s(\mu)$ rapidly increases at $\mu \leq m_s$ , so that
the perturbative result $R^{(n)}(\mu)$, (5.5), becomes unreliable in
such a region.

The relative rations among light quark masses at $\mu=1$ GeV
are fairly reliable, while
the absolute values $m_q(1 {\rm GeV})$ are still controversial.
The relative ratios of light quark masses to heavy quark masses
may be somewhat changed in future.

In this paper, we have evaluated $m_q(\mu)$ only for energy scales $\mu$
which are below the electroweak symmetry breaking energy scale $\Lambda_W$.
Running quark mass values at such an extremely high energy scale far
from $\Lambda_W$ will be given elsewhere.

\vglue.5in

\begin{center}
{\large\bf Acknowledgments}\\[.1in]
\end{center}

This work was first provided as a private memorandum for
mass-matrix-model-building.
The author is indebted to Prof.~M.~Bando,
for encouraging suggestion of making the
private memorandum into a review article for convenience of
mass-matrix model-builders.
The author is very grateful to Prof.~M.~Tanimoto for reading the manuscript
and helpful comments on prescription of running quark masses.
He also thank Dr.~K.~Teshima for consulting treatment of QCD parameter
$\Lambda_{\overline{MS}}^{(n)}$, Prof.~K.~Inoue for comment on
evolution of Yukawa couplings and mass-dependent renormalization
equation, Dr.~P.~Ball for valuable comments on
$\Lambda_{\overline{MS}}$, and Dr.~L.~Surguladze for kindly pointing out
an erratum in the original manuscript.
This work was supported by the Grant-in-Aid for Scientific Research,
Ministry of Education, Science and Culture, Japan (No.06640407).

\vglue.3in
\newcounter{0000}
\centerline{\bf References}
\begin{list}
{[~\arabic{0000}~]}{\usecounter{0000}
\labelwidth=1cm\labelsep=.4cm\setlength{\leftmargin=1.7cm}
{\rightmargin=.4cm}}
\item J.~Gasser and H.~Leutwyler, Phys.~Rep. {\bf 87}, 77 (1982).
\item C.~A.~Dominguez and E.~de Rafael, Ann.~Phys. {\bf 174}, 372 (1987).
\item  S.~Narison, Phys.~Lett. {\bf B216} , 191 (1989).
\item  J.~F.~Donoghue and B.~R.~Holstein, Phys.~Rev.~Lett.
{\bf 69}, 3444 (1992).
\item  C.~A.~Dominguez and N.~Paver, Phys.~Lett. {\bf B293}, 197 (1992).
\item  CDF Collaboration, F.~Abe $et$ $al$., Phys.~Rev.~Lett.
{\bf 73}, 225 (1994).
\item  Particle data group, L.~Montanet $et$  $al$., Phys.~Rev. {\bf D50},
1173 (1994).
\item  N.~Gray, D.~J.~Broadhurt, W.~Grafe and K.~Schilcher, Z.~Phys.
{\bf C48}, 673 (1990).
\item L.~R.~Surguladze, Report No.~OITS 543 (1994) (hep-ph 9405325),
to be published in Phys.~Lett. {\bf B}.
\item  S.~Titard and F.~J.~Yndur\'{a}in, Phys.~Rev. {\bf D49}, 6007 (1994).
\item  H.~S.~Chen and G.~J.~Zhou, Phys.~Lett. {\bf B331}, 437 (1994).
\end{list}

\end{document}